\documentclass[aps,prl,twocolumn,tightenlines,nofootinbib,superscriptaddress]{revtex4-1}
\usepackage{epsfig}
\usepackage{graphicx}
\usepackage{psfrag}
\usepackage{amsmath,amssymb,psfrag,slashed,graphicx}
\usepackage{colordvi}
\usepackage{color}
\usepackage{url}
\usepackage{slashed}
\usepackage{color}
\usepackage[normalem]{ulem}
\usepackage[colorlinks,urlcolor=blue]{hyperref}

\begin{document}

\title{Quantum Anomalous Energy Effects on the Nucleon Mass}

\author{Xiangdong Ji}
\email{xji@umd.edu}
\affiliation{Center for Nuclear Femtography, SURA, 1201 New York Ave. NW, Washington, DC 20005, USA}
\affiliation{Department of Physics, University of Maryland, College Park, MD 20742, USA}

\author{Yizhuang Liu}
\email{yizhuang.liu@sjtu.edu.cn}
\affiliation{Tsung-Dao Lee Institute, Shanghai Jiao Tong University, Shanghai, 200240, China}
\affiliation{Institute fur Theoretische Physik, Universitat Regensburg, D-93040 Regensburg, Germany}
\date{\today}

\begin{abstract}

Apart from the quark and gluon kinetic and potential energies,
the nucleon mass includes a novel energy of pure quantum origin
resulting from anomalous breaking of scale symmetry. We demonstrate the
effects of this quantum anomalous energy (QAE) in QED, as well as in a
toy 1+1 dimensional non-linear sigma model where it
contributes non-perturbatively, in a way resembling the Higgs mechanism
for the masses of matter particles in electro-weak theory. The QAE contribution
to the nucleon mass can be explained using a similar mechanism, in terms of
a dynamical response of the gluonic scalar field through Higgs-like
couplings between the nucleon and scalar resonances. In addition, the QAE
sets the scale for other energies in the nucleon through a relativistic
virial theorem, and contributes a negative pressure to confine the
colored quarks.

\end{abstract}

\maketitle

{\it Introduction}~~~Mass is one of the most important and fundamental
properties of a physical system. For macroscopic ones, mass is simply
the sum of individual parts. The additive rule works to a high degree of
accuracy up until atomic nuclei, for which the masses are the sum of individual nucleon's
(proton and neutron) subtracting the binding energy effects, \'a la Einstein.
Physically, the binding energy takes into account the quantum-mechanical
average of the nucleons' kinetic and interacting potential energies among them
(see for example~\cite{Gandolfi:2020pbj}).
In the electroweak theory, the origin of the masses of elementary particles has a different paradigm: They arise from these particles' interactions with the Higgs potential, which acquires a vacuum condensate after the well-known spontaneous electroweak symmetry breaking, or the Higgs mechanism~\cite{Peskin:1995ev}.

The nucleon mass combines the Higgs mechanism for the quarks with inner workings of quantum chromodynamics (QCD), the fundamental theory of strong interactions. In numerical simulations of QCD on the lattice, the nucleon mass along with masses of other hadrons has been routinely calculated to a very high degree of accuracy~\cite{Durr:2008zz,Fodor:2012gf,Borsanyi:2014jba,Walker-Loud:2018gvh}.
However, to unravel its physical content, one has to analyze the sources of QCD energy according to the famous equation $M=E/c^2$. Since
the original work by one of us~\cite{Ji:1994av,Ji:1995sv}, several papers which calculated the QCD Hamiltonian operator's matrix elements
on the lattice have appeared in the literature ~\cite{Chen:2005mg,Abdel-Rehim:2016won,Alexandrou:2017oeh,Yang:2018nqn}.
Alternative structures of the QCD Hamiltonian have also been proposed and analyzed~\cite{Rothe:1995hu,Lorce:2017xzd,Hatta:2018sqd,Metz:2020vxd}.
Proposals have been made to measure the anomalous gluon matrix element
through heavy-quarkonium electroproduction on the nucleon at Jefferson Lab
and future Electron-Ion Collider~\cite{Kharzeev:1998bz,Hatta:2018ina,Meziani:2020oks}.

The goal of this paper is to further establish the existence and
physical significance of the QCD trace anomaly contribution to
the nucleon mass~\cite{Ji:1994av}.
Although the connection between the trace anomaly and the nucleon mass as well as
the related low-energy theorems have been well studied in the
literature~\cite{Collins:1976yq,Shifman:1978zn}, the anomaly contribution
to the QCD energy itself is a less familiar concept. We consider
the role of this quantum anomalous energy (QAE) in quantum
electrodynamics (QED) where perturbation theory is well-established.
We argue that the QAE holds the key to the QCD nucleon mass generation in two important ways:
First, it sets the scale for the quark and gluon kinetic and potential
energies through a relativistic version of virial theorem. Second,
its contribution to the mass is non-perturbative and bears analogy to the
Higgs mechanism for the masses of elementary matter fields.
We use a simple model, large-$N$ $1+1$ non-linear sigma model,
to demonstrate the second point, and relate the QAE
contribution to the Higgs-like couplings of the scalar resonances
to the nucleon. Finally, the QAE contributes a negative pressure to
confine the colored quarks as in the well-known
MIT bag model~\cite{Chodos:1974je}.

\vspace{5pt}

{\it Scalar and tensor energy and relativistic virial theorem}~~ In classical mechanics,
the virial theorem provides an equation that relates the average over time of the total kinetic energy of a stable system of discrete particles, bound by potential forces, with that of the total potential energy of the system. In quantum field theories (QFTs), a similar relation exists between the matrix elements of the scalar and tensor parts of the Hamiltonian operator, $H=\int d^3 xT^{00}(x)$, where
$T^{\mu\nu}$ is the symmetric energy-momentum tensor. Since any symmetric
second-order tensor can be decomposed into irreducible representations $(1,1)+(0,0)$ of the Lorentz group
$ T^{\mu\nu} = {\bar T}^{\mu\nu} +  {\hat T}^{\mu\nu} $, we can correspondingly
decompose the Hamiltonian into the tensor and scalar parts,
\begin{equation}
H=H_T + H_S \ .
\end{equation}
$H$ is a conserved charge and thus ultra-violet (UV) renormalization scale-independent.
The separate parts, $H_T$ and $H_{S}$, are UV scale-independent as well, a consequence
of Lorentz symmetry. The space-time symmetry further dictates a
relation between average of the tensor and scalar
energies in any stationary state ($\vec{P}=0$)~\cite{Ji:1994av},
\begin{equation}
    E_T = (d-1) \, E_S
\end{equation}
where $E_{T,S}=\langle H_{T,S}\rangle$ is a quantum average in the zero momentum hadron state, and $d=4$
is the space-time dimension. This has been referred to as a virial theorem
in QFT~\cite{Ji:1994av}.

The above relation has to do with the virial theorem in classical physics and
may be seen heuristically as follows. In classical mechanics, the virial theorem relates the average kinetic energy $\langle K\rangle$ and the total potential energy $\langle V\rangle$ of a system, with the exact coefficient depending on the scaling property of the potential under scale transformation $\vec{r} \to \lambda\vec{r}$,
where $\lambda$ is a scale factor. In QFT, the behavior of a system under scale transformation
depends on the dilatation current $j^\mu_D= x^\alpha T^\mu_{\,\,\, \alpha}$, which has a divergence $ \partial_\mu j^\mu_D
= T^\alpha_{\,\,\,\alpha} \sim dH_S$. {\it Thus the scalar energy density, $H_S$, comes entirely
from scale-breaking effects.} In gauge theories such as QED and
QCD, $H_T$ reads
\begin{align}
H_{T}=\int d^d \vec{x} \bigg[\bar \psi(i\vec{\gamma}\cdot \vec{D}+\frac{d-1}{d}m) \psi+\frac{1}{2}\left(\vec{E}^2+\vec{B}^2\right)\bigg] \ .
\end{align}
where $\psi$ is a fermion field with $m$ as its mass, $\vec{E}$ and $\vec{B}$ are the electric
and magnetic fields, $D$ is a covariant derivative.  Thus the tensor energy $E_T$ includes the
familiar kinetic and potential energies of particles and fields.
Eq.(2) can be interpreted as the scalar energy $E_S$ sets a scale for the tensor energy $E_T$ which grows linearly with space dimension $d-1$.

The most familiar scale-breaking effects is the mass terms such as $m\bar\psi\psi$ for Dirac fields $\psi$
or $\mu^2 \phi^2$ for scalar field $\phi$. In the case of QED, the electron
mass is the only mechanical scale which, together with the dimensionless coupling $\alpha\sim 1/137$,
sets the energy scale for atomic physics, chemistry and biology. In cases like 1+1 dimensional QCD,
the dimensional coupling introduces scalar energy as well~\cite{Ji:2020bby}. An important mechanism
to generate a new source of scale-breaking is through the condensates of scalar fields in the
ground states of QFTs. This mechanism has been used to provide masses of elementary particles in the standard model
(the Higgs mechanism)~\cite{Peskin:1995ev}, and energy needed for the inflationary universe as well as a
mechanism for cosmic dark energy~\cite{Perkins:2003pp}.

It is easy to check that in the non-relativistic limit, the above equation for the positronium system
in 3+1 quantum electrodynamics (QED) reduces to the known virial theorem in quantum mechanics. Indeed in Coulomb gauge $\nabla \cdot \vec{A}=0$, and in the leading component of the non-relativistic positronium state, one can show that Eq. (2)
\begin{align}
   2\langle K \rangle+\langle V\rangle=0
\end{align}
which is nothing but the non-relativistic virial theorem. It is
easy to check that this statement is actually gauge invariant.

\vspace{5pt}

{\it Quantum anomalous energy and its perturbative role in QED}~~~
The QAE is a novel scale-breaking energy source arising from short-distance
quantum fluctuations. The UV physics caused by these quantum effects
cannot be completely removed but leave a trace through running of the coupling
constants and associated composite scalar fields as an anomalous contribution to Hamiltonian
which represents those ``residue memories'' of the microscopic world.

To see this new energy emerging from scaling
breaking, we consider a re-scaling in time direction, $t\rightarrow (1+\delta \lambda) t$. In this case, the cutoff in the temporal direction needs to change accordingly and results in asymmetric cutoff theory. The anomalous contribution to the Hamiltonian can be derived by studying the response of the system under the re-scaling, $S\rightarrow S+\delta\lambda \int dt H$~\cite{DiFrancesco:1997nk}. For example, for the classical gauge theory, in terms of the time re-scaled field variables $A_{\mu}'(t',\vec{x})=A_{\mu}(t,\vec{x})$, the action transforms to
\begin{align} \label{eq:assy}
\frac{1}{2g_0^2}\int d^4x' \bigg(-\frac{1}{1+\delta \lambda} ({\vec B}')^2+(1+\delta \lambda)(\vec{E}')^2\bigg) \ .
\end{align}
The contribution at order $\delta \lambda$ is nothing but the Hamiltonian $H=\int d^3\vec{x}\frac{1}{2g_0^2}(\vec{E}^2+\vec{B}^2)$ integrated over time, consistent with the general principle above.  However, in quantum theory and in the presence of a cutoff in the temporal direction, such as a lattice cutoff, the $g_0$ in Eq.~(\ref{eq:assy}) must depends on $\delta \lambda$ in order to be equivalent to the time translated theory~\cite{Karsch:1982ve}. Therefore, the derivative of $g_0$  with respect to $\delta \lambda$ will generates anomalous terms in addition to the canonical energy, which has been shown to be exactly the anomalous contribution
~\cite{Rothe:1995av}
\begin{align}\label{eq:anoaH}
H_a=\frac{1}{4}\int d^3 \vec{x}\bigg(\frac{\beta(g_0)}{2g_0}F^2+m_0\gamma_m\bar\psi \psi \bigg) \ .
\end{align}
where $F^{\mu\nu}$ is the gauge field strength with coupling $g_0$, which
arises from logarithmic running of couplings $m_0$ and $g_0$ with the UV
cut-off scale, as reflected in the beta function $\beta(g_0)$ and the mass anomalous dimension $\gamma_m$. This agrees with the original mass decomposition in Ref.~\cite{Ji:1994av} in which the role of the trace anomaly is explicit. The mass decomposition in Ref.~\cite{Metz:2020vxd}, although technically correct in dimensional regularization, missed
the key role that the anomalous term plays. We call the expectation value of $H_a$ in a state, $E_a$, the quantum anomalous energy.

There is an anomalous energy contribution to the QED Hamiltonian.
However, due to perturbative nature of the theory, this anomalous energy does not bring in
any new scale, and its contribution is embedded in covariant perturbation theory and
is proportional to the electron mass. The electron
pole mass $m_e$ at one-loop level receives a contribution from the mass anomalous dimension term,
$ \langle e|H_a|e\rangle=\frac{3\alpha m_e}{8\pi}$. In a external field ${\cal A}$, the anomalous energy contains a mixing term between the external field and the radiative field. As such, this will contributes to Lamb shift for hydrogen atom which is at order $\alpha$ in radiative field.  A calculation shows that to leading order in radiative correction, the anomalous part leads to the contribution
\begin{align}\label{eq:tracelamb}
\langle    n,j|H_a|n,j\rangle=\frac{\alpha^2}{6\pi}\int d^3\vec{x}\frac{u_{n,j}^{\dagger}(\vec{x})u_{n,j}(\vec{x})}{|\vec{x}|}+\frac{3\alpha}{8\pi}E_{n,j} \ ,
\end{align}
for the energy level $n,j$, where $n$ is the radial quantum number and $j$ is the total spin. The first term is the photonic contribution while the second terms is the fermionic contribution. Here $u_{n,j}(\vec{x})$ is the quantum-mechanical wave function that solves the Dirac equation in a static Coulumb field, and $E_{n,j}$ is the bound state energy. In the non-relativistic limit, Eq.~(\ref{eq:tracelamb}) can be further expanded in $\alpha$ and contains contributions at ${\cal O}(\alpha)$, ${\cal O}(\alpha^3)$ and ${\cal O}(\alpha^5)$. The ${\cal O}(\alpha)$ and ${\cal O}(\alpha^3)$ contributions will be cancelled by other terms,
while the contribution at ${\cal O}(\alpha^5)$ reads
\begin{align}
\langle    n,j|H_a|n,j\rangle^{(5)}=-\frac{7m_e\alpha^5}{24\pi n^4}\left(\frac{3}{8}-\frac{1}{2j+1}\right) \ .
\end{align}
This contributes to the famous Lamb shift at ${\cal O}(\alpha^5)$.

\vspace{5pt}

{\it Non-perturbative QAE and dynamical Higgs mechanism in 1+1 nonlinear sigma model}~~~
In more interesting cases, the QAE will generate a non-perturbative contribution characterized with
a new mass scale (dimensional transmutation~\cite{Coleman:1974hr}). On the other hand,
the anomalous scalar field can be considered as a dynamical one, and the QAE contribution to the
mass comes from its dynamical response to the matter, in analogy to the Higgs
mechanism for fermion masses in the standard model.

\begin{figure}[t]
\includegraphics[width=0.4\columnwidth]{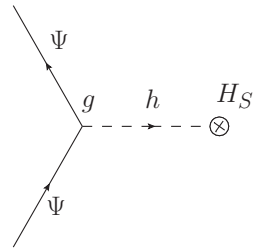}
\caption{Dynamical response of the scalar Hamiltonian $H_S$
in the presence of the fermion $\Psi$, generating a contribution to the fermion mass  The dotted line
represents the dynamical Higgs particles $h$ and the crossed circle denotes the scalar Hamiltonian linear in $h$. The coupling $g$ between the Higgs field and the fermion is proportional to fermion mass. }
\label{fig:higgs}
\end{figure}

To see this analogy, lets first review the Higgs mechanism for matter particles
with a simplified scalar field  $\Phi$ with potential $V(\Phi)=-\frac{\mu^2}{2}\Phi^2+\frac{\lambda}{4!}\Phi^4$ which couples to a massless fermion $\Psi$ through Yukawa interaction $-g\bar \Psi \Psi \Phi$. At tree level, the scalar $\Phi$ develops a condensate at $\langle \Phi\rangle^2=\frac{6\mu^2}{\lambda}$, which gives the fermion mass $m_\Psi = g\langle \Phi\rangle$.
On the other hand, one can show by equation of motion that, in terms of the dynamical $h=\Phi-\langle \Phi\rangle$, the scalar part of the Hamiltonian at lowest order is linear in $h$ and equals to $H_S=-\frac{1}{2}\int d^4x \mu^2\langle \Phi\rangle h$. In the presence of the fermion $\Psi$, the quantum field $h$ generates a response which
contribute to the fermion mass through the intermediate Higgs particle:  $\langle \Psi|H_S|\Psi\rangle = (-g)f_s/m_h^2 = (1/4)m_\Psi$, where $f_s=-\frac{1}{2}\mu^2\langle \Phi\rangle$ is a scalar decay constant and $m_h=\sqrt{2}\mu$ is the Higgs mass. The $\frac{1}{m_h^2}$ is due to the zero-momentum propagator of the Higgs field. Therefore, the scalar part of the Hamiltonian contributes 1/4 of the fermion mass through the dynamical Higgs. See Fig.~\ref{fig:higgs} for a depiction of the mechanism. The example demonstrates that the scalar part of the mass of the fermions can also be measured by the response of the fluctuating part of the scalar field in the presence of the matter.

In the above model, the Higgs scalar field is introduced at the Lagrangian level ``by hand''. However, in case that the scalar field emerges dynamically, the scenario remains qualitatively the same.  We use the example of the 1+1 dimensional nonlinear sigma model in the large $N$ limit to demonstrate it. The model consists of an $N$ component scalar field $\pi=(\pi^1,...\pi^N)$ which lives in the unit $N-1$ sphere $\sum_{a=i}^N\pi^i\pi^i=1$ with the action
\begin{align}\label{eq:no-linea}
S=\frac{1}{2g_0^2}\int d^2x \sum_{i=1}^N(\partial_{\mu} \pi^i)(\partial_{\mu} \pi^i) \ .
\end{align}
Here, $g_0$ is a dimensionless coupling. The theory is scale-invariant at classical level, but
the scale symmetry is broken by a UV cut-off at quantum level and the
coupling becomes scale dependent. The model can be analytically solved in the large-$N$ limit where the coupling $\lambda_0 = g_0^2N$ stays finite~\cite{Novikov:1984ac,Shifman:2012zz}. The constraint can be removed by adding an auxiliary term to the action,
\begin{align}
S_{ \Sigma}=\frac{i}{2g_0^2}\int d^2x  \Sigma(\sum_{i=1}^N \pi^i\pi^i-1) \ .
\end{align}
By integrating out the $\pi^i$ fields, an effective potential for $ \Sigma$ can been generated, which
has a saddle point $\langle  \Sigma \rangle=-im^2$ and introduces a new mass scale.
The condensate $m^2$ satisfies the gap equation
\begin{align}\label{eq:gapsigma}
\frac{1}{g_0^2N}=\int \frac{d^2k}{(2\pi)^2}\frac{1}{k^2+m^2} \sim \frac{1}{4\pi}\ln \frac{\Lambda_{\rm UV}^2}{m^2} \ ,
\end{align}
which determines the bare $\lambda_0$ as a function of $m$ and the UV cutoff $\Lambda_{\rm UV}$.
Therefore, the theory is asymptotically free with the beta function $\beta(g_0)=-\frac{Ng_0^3}{4\pi}$. Furthermore, the vacuum condensate for $\Sigma$ generates a mass $m$ for $\pi^i$, and the spectrum consists of $N$ massive scalars $\pi^a$ with equal mass.

The anomalous part of the Hamiltonian is
\begin{align}\label{eq:anoapi}
H_a=-\frac{\beta(g_0)}{2g_0}\int dx^1 \sum_{i=1}^N(\partial_{\mu} \pi^i)(\partial_{\mu} \pi^i) \ .
\end{align}
Although in the limit of $g_0 \to 0$, $\frac{\beta(g_0)}{2g_0}=-\frac{Ng_0^2}{4\pi}$ is proportional to $g_0^2$,
$H_a$ does not vanish.  The operator $(\partial_{\mu} \pi^i)^2$ receive quantum fluctuations from the loop diagram proportional to
\begin{align}
\int \frac{d^2k}{(2\pi)^2}\frac{k^2}{(k^2+m^2)^2} \ ,
\end{align}
which diverges logarithmically. It can be shown that the contribution to the $\pi^i$ mass is always 1/2
independent of regularization scheme,
\begin{equation}
               \langle \pi^i|H_a|\pi^i \rangle = \frac{m}{2}
\end{equation}
which is consistent with the virial theorem in Eq. (2).

The QAE contribution to the meson mass
can be explained in term of a dynamical Higgs mechanism as follows.
Using the equation of motion, the anomalous Hamiltonian can also be re-written in terms of the auxiliary scalar
\begin{align}
H_a=-\frac{iNm^2}{8\pi}\int dx^1\sigma \ ,
\end{align}
where the dimensionless scalar $\sigma= (\Sigma-\langle \Sigma \rangle)/m^2$ contains the quantum fluctuation part. This is similar to the Higgs example above, in that the scalar part of the Hamiltonian is linear in the sigma field.
Its contribution to the pion mass is determined by $\langle \pi^i|\sigma|\pi^i\rangle$.
By using the $\pi\pi\sigma$ vertices in Eq.~(10), and the dominance of the
zero-momentum $\sigma$ propagator $\langle \sigma(0)\sigma(0) \rangle=8\pi /(Nm^2)$
in the intermediate state, the response of the scalar $\sigma$ to
$\pi^i$ state exactly makes $H_a$ contributing $\frac{1}{2}$ of the $\pi^i$ mass. We shall mention that the propagator of $\sigma$~\cite{Shifman:2012zz} contains only a cut starting at the two-$\pi$ threshold $p^2=4m^2$ but no poles, unlike the Higgs field $h$ in the previous example. Nevertheless, the zero-momentum propagator of $\sigma$ contributes to the average of the anomalous Hamiltonian exactly the same way as the zero-momentum propagator of the Higgs field $h$.

\vspace{5pt}

{\it Dynamical scalar and QAE contribution to the nucleon mass and pressure}~~~For simplicity,
we consider the limiting case of massless up and down quarks. The anomalous
Hamiltonian comes entirely from the gluon composite scalar, $H_a = \int d^3\vec{x} \Phi(x)$,
where $\Phi(x) = \beta(g)/(8g)F^{\mu\nu}F_{\mu\nu}(x)$. As in the non-linear sigma model,
its contribution to the nucleon mass can be seen as a form of dynamical Higgs-mechanism,
which is consistent with that the Higgs and confining phases of matter-coupled gauge theory are smoothly connected~\cite{Fradkin:1978dv,Banks:1979fi}.

It is useful to recall that for the infinite-heavy $\bar Q Q$ state separated by $r$ in pure gauge theory, it has been shown~\cite{Rothe:1995hu,Rothe:1995av} that the non-perturbative
contribution of $H_a$ to the static potential is $\frac{1}{4}(V(r)+rV'(r))$. At large $r$ where the confinement potential
dominate $V(r)\sim \sigma r$, the anomalous contribution is exactly one half of the confinement potential.

The scalar field $\Phi(x)$ has a vacuum condensate $\Phi_0=\langle 0|\Phi|0\rangle$~\cite{Shifman:1978bx,Shifman:1978by}.
However, in the presence of the nucleon, the quantum response is measured by
\begin{equation}
      \phi (x) = \Phi(x) - \Phi_0 \ ,
\end{equation}
which is a dynamical version of the MIT bag-model constant $B$~\cite{Chodos:1974je}.
Its contribution to the nucleon mass can be seen as the response
of the scalar field to the nucleon source,
\begin{equation}
    E_a = \langle \phi\rangle_N = \langle N|\phi(x)|N\rangle \ ,
\end{equation}
where the nucleon state is normalized as $\langle N|N\rangle=(2\pi)^3\delta^3(0)$.
If $\phi(x)$ is a static constant $B$ inside the nucleon, $E_a$ will be of order $
BV$, where $V$ is the effective volume in which the valence quarks
are present.

The static response of the composite gluon scalar $\phi$ in the nucleon state
can be measured in the electro-production of heavy quarkonium on the proton~\cite{Kharzeev:1995ij,Kharzeev:1998bz,Brodsky:2000zc,Hatta:2019lxo,Wang:2019mza,Du:2020bqj,Zeng:2020coc} or leptoproduction of heavy quarkonium at large photon virtuality~\cite{Boussarie:2020vmu}. The color dipole from the quarkonium will be an effective probe of the $F^2$. This also provides a direct determination of the QAE contribution to the mass.

\begin{figure}[t]
\includegraphics[width=0.4\columnwidth]{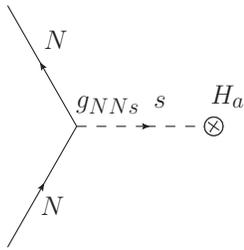}
\caption{Quantum anomalous energy contribution to the nucleon mass seen as
a dynamical response of the anomalous scalar field in the presence of the nucleon. The dotted line
represents the intermediate scalar particles with couplings $g_{NNs}$ proportional to the nucleon mass, which is
dominated by a single Higgs particle in the Higgs mechanism.}
\label{fig:dispersion}
\end{figure}

An interesting mechanism for its physics is to consider a dynamical
response of the $\phi$ in the presence of the nucleon through a tower of scalar $0^{++}$
spectral states, as in the Higgs model.  Assume an effective
coupling between the nucleon and scalar $g_{NN\phi}\bar NN \phi$,
the QAE contribution to the mass can be related to the scalar field response function,
\begin{equation}\label{eq:disper1}
    \langle N|\phi| N\rangle =ig_{NN\phi}\langle \phi(0) \phi(0) \rangle
\end{equation}
where $\langle \phi(0) \phi(0) \rangle$ is the zero-momentum propagator of the scalar field $\phi$.
If the propagator is dominated by a series of scalar resonances,
or $\langle \phi(0) \phi(0) \rangle=\sum_s \frac{if_s^2}{-m_s^2}$, one has
\begin{align}\label{eq:disper2}
 \langle N|\phi| N\rangle =\sum_s \frac{g_{NNs}f_s}{m_s^2} \ .
\end{align}
Here $m_s$ is the mass of the scalar resonances, $f_s=\langle s|\phi|0\rangle$ is the decay
constant and $g_{NNs}\equiv g_{NN\phi}f_s$ is the coupling of the nucleon to the scalars.
See Fig.~\ref{fig:dispersion} for a depiction.

One might assume the dominance of the lowest mass scalar glueball-like state, generically called $\sigma$, for the above equation. If the coupling constant $g_{NNs}$ can be extracted through experiment, one can perform a consistency check on the $\sigma$ dominance picture by combining the glueball masses and the decay constants extracted from lattice QCD calculations~\cite{Morningstar:1999rf,Chen:2005mg}. In fact, for the lowest glueball state $\sigma$, low-energy theorem predicts~\cite{Ellis:1984jv} that $f_{\sigma}=m_\sigma\sqrt{|\Phi_0|}$. Given this relation and assuming the sigma dominance, we predict that $g_{NN\sigma}=\frac{m_Nm_\sigma}{4\sqrt{|\Phi_0 |}}$ in the chiral limit. This then exactly corresponds to the Higgs model mentioned earlier. Of course, the QCD reality
is in between the simple Higgs and the 1+1 sigma models. However, the coupling
between the nucleon (or any other hadrons) and with the scalars must be proportional to the
to the mass, same as in the Higgs case which has been tested recently at LHC~\cite{Weinberg:1967tq,Sirunyan:2018koj,Aad:2019mbh}.

For pion state, it has been shown~\cite{Ji:1994av} that $H_a$ contributes to $\frac{1}{8}$ of the total pion mass. Assuming the $\sigma$ dominance, the effective coupling between the pion and the scalar glueball $g_{\sigma \pi\pi}$ is again proportional to the pion mass, in consistent with a dynamical Higgs effect.

Finally, the $\phi$ contributes a negative mechanical pressure
to the trace part of the energy-momentum tensor~\cite{Ji:1995sv}, just like the cosmological constant does
in Einstein's gravity theory~\cite{Perkins:2003pp}. The physics of this has been well explored
in the context of MIT bag model~\cite{Chodos:1974je}.
Its contribution confines the colored quarks and cancels the positive quarks and gluon
contributions, which are measurable through deeply virtual Compton scattering~\cite{Ji:1996nm,Polyakov:2002yz,Burkert:2018bqq}.

To conclude, the mass of the nucleon contains a quantum anomalous contribution
which sets the scale for other type of contributions such as quark and gluon kinetic and
potential energies. This contribution has a physical mechanism similar to the Higgs model,
with a dynamical scalar field generating a response, having the characteristic feature
that the coupling to the scalars is proportional to the fermion mass. Furthermore, it contributes
a negative pressure to confine the colored quarks.

In preparation of the paper, there appeared another calculation of anomalous energy contribution to hydrogen atom mass~\cite{Sun:2020ksc}. Their result differs from ours by a factor of $2$.

{\it Acknowledgment.}---We thank J. C. Peng and Z. Meziani for discussions related to the proton mass. This material is supported by the U.S. Department of Energy, Office of Science, Office of Nuclear Physics, under contract number DE-SC0020682.

\bibliographystyle{apsrev4-1}
\bibliography{bibliography}

\end{document}